\documentclass[]{spie}
\usepackage{amsfonts}  
\usepackage[]{graphicx}
\usepackage{subfigure}

\title{Spin photonics and spin-photonic devices with dielectric metasurfaces}

\author{Yachao Liu, Shizhen Chen, Yougang Ke, Xinxing Zhou, Hailu Luo$^{*}$,
and Shuangchun Wen\skiplinehalf \supit{}Laboratory for Spin
Photonics, School of Physics and Electronics, Hunan University,
Changsha 410082, China\nonumber}

\authorinfo{\\$^{*}$Hailu Luo email address: hailuluo@hnu.edu.cn}

\begin{document}
  \maketitle

\begin{abstract}
Dielectric metasurfaces with spatially varying birefringence and
high transmission efficiency can exhibit exceptional abilities for
controlling the photonic spin states. We present here some of our
works on spin photonics and spin-photonic devices with metasurfaces.
We develop a hybrid-order Poincar\'{e} sphere to describe the
evolution of spin states of wave propagation in the metasurface.
Both the Berry curvature and the Pancharatnam-Berry phase on the
hybrid-order Poincar\'{e} sphere are demonstrated to be proportional
to the variation of total angular momentum. Based on the
spin-dependent property of Pancharatnam-Berry phase, we find that
the photonic spin Hall effect can be observed when breaking the
rotational symmetry of metasurfaces. Moreover, we show that the
dielectric metasurfaces can provide great flexibility in the design
of novel spin-photonic devices such as spin filter and
spin-dependent beam splitter.
\end{abstract}



\section{INTRODUCTION}
Spin provides a reliable degree of freedom for light control and
optical data
storage~\cite{Onoda2004,Yin2013,Shitrit2013,Bliokh2015}. To employ
this degree of freedom, an effective method for spin-orbital
coupling is required in the light-matter interaction. This coupling
will lead to a featured spin dependent response in optical materials
or in the emerging fields. For particular responses, it can be
applied for different uses such as the spin-controlled plasmonic
transmission and the spin dependent beam
steering~\cite{Lin2013,Huang2013}. To design the electromagnetic
response, many material systems have been studied in recent
years~\cite{Litchinitser2012,Kildishev2013}. The phase gradient
metasurfaces have been verified as a giant photonic spin Hall effect
(SHE) generator~\cite{Yin2013}. Quantum dots and nitrogen-vacancy
centers in diamond are believed as the candidates for quantum
interface between stationary and flying qubits~\cite{Gao2012}. Also
such as the plasmonic structures be designed for the near-field
vortex mode generation and the polarization sensitive
focusing~\cite{Bliokh2008b,Gorodetski2008}.

Metasurfaces can artificially design the spin-dependent or the
orbital-relevant responses of the emerging field, thus is a
promising route to mould the
light~\cite{Yin2013,Shitrit2013,Kildishev2013,Yu2013,Meinzer2014}.
The dispersion engineering of metasurfaces can be realized by
changing the material compositions, the pattern of local components,
and the arrangement of these local
patterns~\cite{Yu2011,Huang2012,Genevet2012}. The flexible designing
and reduced dimension of metasurfaces make it potentially a
wonderful candidate for the on-chip optics and the integrated
optics~\cite{Sun2012a,Ni2013,Liu2013}. To achieve a spin-dependent
behavior, the comprising sub-wavelength structures must be
polarization (spin) sensitive. The polarization (spin) related
evolving process of light is affirmatively to impose a geometric
phase into the light, which is usually called Pancharatnam-Berry
(PB) phase~\cite{Bliokh2008b,Bliokh2008a,Ding2015}. There is
therefore a strong motivation to peer into the spin-dependent PB
phase for exploring the applications of metasurface in spin
photonics.

In the paper, we reviewed some of our works on the spin photonics
and spin photonic devices base on the dielectric metasurfaces. The
Poincar\'{e} sphere is an intuitive tool to understand the
polarization (spin) and spatial optical mode evolving
process~\cite{Poincare1892,Padgett1999,Milione2011,Holleczek2011}.
We developed a hybrid-order Poincar\'{e} sphere to describe the
evolution of polarization states of wave propagation in
inhomogeneous anisotropic media~\cite{Yi2015}. And the dielectric
metasurfaces are applied to show examples of this
evolutions~\cite{Liu2014,Yi2014}. Furthermore, the dielectric
metasurfaces can dramatically reduce the losses induced in the
propagation~\cite{Lin2014}, which limited the practical application
of metamaterials since its advent. Therefore, the dielectric
metasurfaces are expected to be alternative in the integrated
optics. We provided some schemes base on the dielectric metasurfaces
to realize the giant photonic SHE generater, mode splitter, and spin
filter. The rest of the paper is organized as follows. In Sec. 2, we
establish a hybrid-order Poincar\'{e} sphere model to describe the
complicate evolutions of polarization and deduce the involved PB
phases. Moreover, we introduce the precondition for the photonic SHE
caused by PB phase in the dielectric metasurfaces. In the Sec. 3,
some spin-photonics devices based on the dielectric metasurfaces are
demonstrated. Finally, a conclusion is given in Sec. 4.

\begin{figure}
\centerline{\includegraphics[width=17cm]{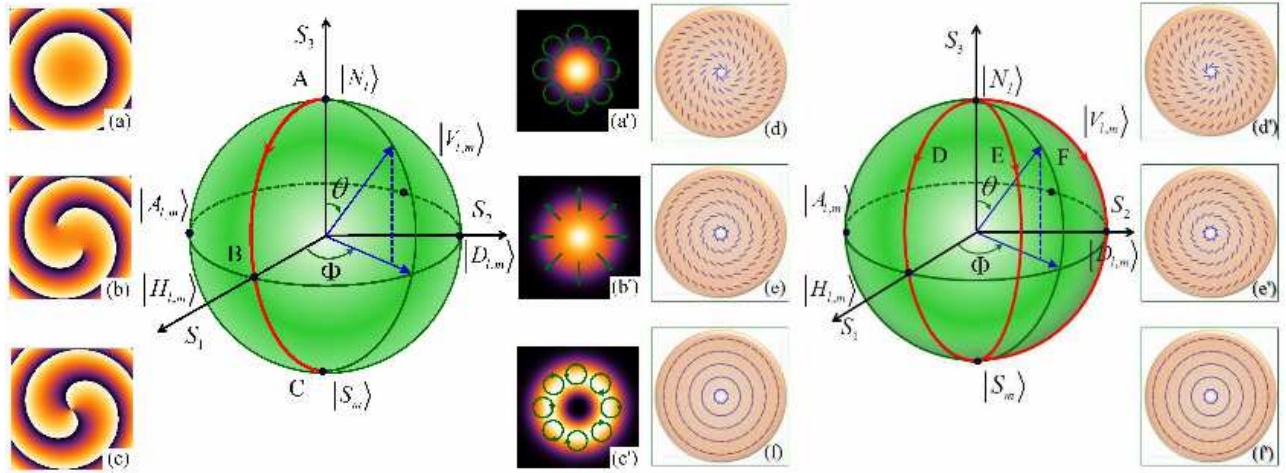}}
\caption{The left schematic illustration shows the evolution of
phase and polarization on hybrid-order Poincar\'{e} sphere. The
sphere is assumed with state $\sigma=+1$ and $l=0$ in the north
pole, and the state $\sigma=-1$ and $m=+2$ in the south pole. Insets
(a)-(c) show the the phase for points A, B, and C, respectively.
Insets (a$'$)-(c$'$) show the polarization state of these points
respectively. The right schematic illustration shows realization of
the evolution along different longitude lines on the hybrid-order
Poincar\'{e} sphere ($l=0$ and $m=-2$). By choosing the initial spin
state $\sigma$ and the initial angle $\alpha_{0}$ of the dielectric
metasurfaces, we can change the evolving longitudes. Insets (d)-(f)
are the evolutions along the longitude lines D, E, and F from the
north pole to south pole, respectively. While, insets (d$'$)-(f$'$)
are the evolutions along the longitude lines D, E, and F from south
pole to north pole. The initial angle of metasurfaces can be
obtained by the relation $\Phi= 2\alpha_{0}-\pi/2$ from the north
pole to south pole, and $\Phi= 2\alpha_{0}+\pi/2$ from the south
pole to north pole~\cite{Yi2015}.}\label{sphere}
\end{figure}

\section{HYBRID-ORDER POINCAR\'{E} SPHERE}
Polarization and phase are two intrinsic features of electromagnetic
waves. To describe the fundamental homogeneous polarization states,
a prominent geometric representation known as the Poincar\'{e}
sphere was proposed in 1892, in which the homogeneous polarization
states are mapped to the surface of a unit sphere through the Stokes
parameters in the sphere's Cartesian
coordinates~\cite{Poincare1892}. Analogue to this geometric
representation of polarization, the orbital Poincar\'{e} sphere was
proposed to describe the spatial mode of beam in recent, in which
each point on the unit spherical surface represents a unique spatial
mode of the same order~\cite{Padgett1999}. Both of these
representations provide a intuitive way to treat the complicated
phenomena with varying states. It is well known that the cyclic
transformation of a light beam will endow the light an additional
geometric phase. Therefore, the Poincar\'{e} sphere representations
are definitely efficient approaches to find these additional phases.

However, the researches of spatial inhomogeneous polarization states
have broken through the limit of general Poincar\'{e} sphere
recently. For example, cylindrical vector beam refers to the
polarization state with cylindrical symmetry~\cite{Zhan2009}, which
attracted significant attentions because of their unique properties
under high-numerical-aperture focusing~\cite{Novotny2001,Dorn2003}.
To demonstrate the transformation of these higher-order solutions of
Maxwell¡¯s equations, the high-order Poincar\'{e} sphere has been
developed in the most recent~\cite{Milione2011,Holleczek2011}. The
north and south poles of the high-order Poincar\'{e} sphere
represent the opposite spin states and orbital states. Any states on
the high-order Poincar\'{e} sphere, can be realized by a
superposition of the two orthogonal states. The same as the cyclic
evolution of fundamental polarization state will lead to an
additional phase, which is named as the PB phase, the higher-order
polarization state evolution will also lead to a higher-order PB
phase~\cite{Milione2012}.

Nevertheless, all the above solutions can not describe the
polarization transforms from a homogeneous state to a higher-order
state. These cases are usually involved in the inhomogeneous
anisotropic media interactions~\cite{Marrucci2006}. To address this
concern, we present a hybrid-order Poincar\'{e} sphere to
demonstrate the more comprehensive evolution of polarization and
phase~\cite{Yi2015}. In the constructed parameter space, all the
states of hybrid-order Poincar\'{e} sphere is given by
\begin{equation}
|\mathbf{\psi(\theta,\Phi)}\rangle=\cos\frac{\theta}{2}|\mathbf{N}_l\rangle
+\sin\frac{\theta}{2}|\mathbf{S}_m\rangle
\exp(+i\sigma\Phi)\label{HOPS},
\end{equation}
where
$|\mathbf{N}_l\rangle=\frac{\sqrt2}{2}(\mathbf{e}_x+i\sigma\mathbf{e}_y)\exp(il\Phi/2)\label{NL}$,
and
$|\mathbf{S}_m\rangle=\frac{\sqrt2}{2}(\mathbf{e}_x-i\sigma\mathbf{e}_y)\exp(im\Phi/2)\label{SM}$
are polarization and phase states represented by the poles. The
coordinates $(\theta,\Phi)$ are respectively the latitude and
longitude on the sphere. The optical spin angular momentum (SAM) of
a photon is $\sigma\hbar$ ($\sigma=\pm1$), $\hbar$ is the reduced
Plank constant; the azimuthal phase factor $\exp(il\Phi)$ is a phase
vortex, the optical orbital angular momentum (OAM) eigenstates of a
photon is $l\hbar/m\hbar$ ($l,m=0,\pm1,\pm2\ldots $), $l/m$ is the
topological charge; and $(\mathbf{e}_x\pm i\sigma\mathbf{e}_y)$ is
right or left circular polarization depending on the spin of photon
$\sigma$. Figure~\ref{sphere} schematically shows the idea of
hybrid-order Poincar\'{e} sphere.

\begin{figure}
\centerline{\includegraphics[width=17cm]{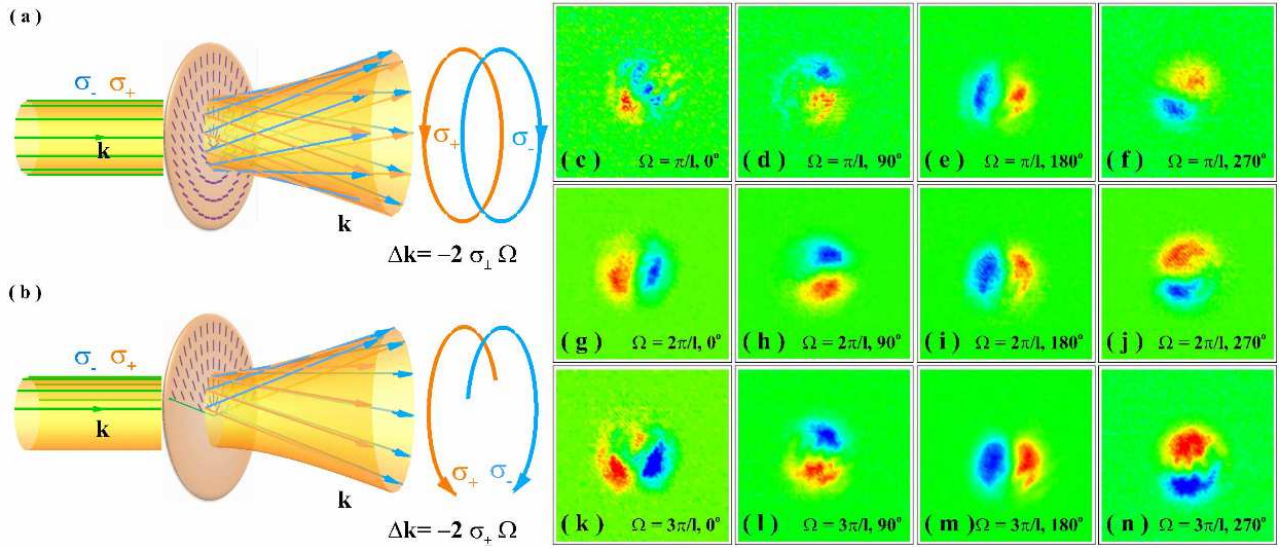}}
\caption{(a)-(b) Contrast between the metasurfaces (a) with
rotational symmetry and (b) with rotational symmetry breaking. The
rotational symmetry can be broken by designing the structure of the
metasurface, which just maintains part of the inner structure (b)
comparing with the metasurface with rotational symmetry (a), or by
displacing the incident beam from the central axis of the
metasurface. The notations $\sigma_{+}$ and $\sigma_{-}$ represent
left- and right-spin states, respectively, and $\Omega$ is rotation
rate of the metasurface. (c)-(n) The S3 results of photonic SHE when
we break the rotational symmetry of PB phase; red and blue represent
the left and right-circular spin states, respectively. The three
rows correspond to the results of three different metasurfaces
measured in four different positions $\Psi= 0^{\circ}, 90^{\circ},
180^{\circ},$ and $270^{\circ}$~\cite{Liu2015}.}\label{split}
\end{figure}

It is worth noting that the hybrid-order Poincar\'{e} sphere has two
special features in comparing with the high-order Poincar\'{e}
sphere: (1) The orbital states on the hybrid-order Poincar\'{e}
sphere should not be confined to have the same values and opposite
signs. As a result, the cylindrical vector vortex beams can be
represented by the equatorial points. Intermediate points between
the poles and equator represent the elliptically polarized vector
vortex beams. (2) Polarization and phase evolution in any
metasurfaces with cylindrically symmetrical optical axis
distribution can be conveniently described by state evolution along
the longitude line on the hybrid-order Poincar\'{e} sphere.
Therefore, the hybrid-order Poincar\'{e} sphere is a predictable way
to access the PB phase involved in the optical transform inside an
inhomogeneous anisotropic medium. The dielectric metasurface is
expected to be a good candidate for realizing the evolution of
polarization states on the hybrid Poincar\'{e} sphere. By correctly
controlling the local orientation and geometrical parameters of the
nanograting, one can achieve any desired polarization distributions
on the hybrid-order Poincar\'{e} sphere, as the right picture of
Fig.~\ref{sphere} showing.

To get the PB phase, we deduced the Berry connection and the Berry
curvature when the states evolve on the surface of the hybrid-order
Poincar\'{e} sphere. The Berry connection can be written as
\begin{equation}
\mathbf{A}=i\langle\psi(\mathbf{R})|\nabla_R|\psi(\mathbf{R})\rangle\label{BerryC}.
\end{equation}
Where $\mathbf{R}=(\rho, \theta, \Phi)$ is the coordinate in
parameter space. After substituting, we find the only nonzero
component of Berry connection in our case:
\begin{equation}
\mathbf{A}_\Phi=-\frac{1}{4\rho\sin\theta}[l(1+\cos\theta)
+(m+2\sigma)(1+\cos\theta)]\label{BerryCPhi}.
\end{equation}
Then we get the Berry curvature $\mathbf{V(R)}$ and PB phase
$\gamma(C)$ in hybrid-order Poincar\'{e} sphere:
\begin{equation}
\mathbf{V(R)}=-\nabla_\mathbf{R}\times\mathbf{A}=\frac{l-(m+2\sigma)}{4\rho^2}
\hat{\rho}\label{BerryCI},
\end{equation}
\begin{equation}
\gamma(C)=-\int\int_{C}d\mathbf{S}\cdot\mathbf{V(R)}=-\frac{l-(m+2\sigma)}{4}\Theta\label{GPII},
\end{equation}
where
$d\mathbf{S}=\rho^2\sin{\theta}d{\rho}d{\theta}d{\Phi}\hat{\rho}$,
$C$ is the route of a cyclic transformation in parameter space
$\mathbf{R}$, $\Theta$ is the surface area on the hybrid-order
Poincac\'{e} sphere enclosed by the circuit $C$.

The PB phase is immanent in the spin-orbit interaction (SOI)
processes with polarization varying, which may lead to the photonic
SHE. However, the precondition for observing this photonic SHE is
still obfuscated. We present that it is necessary to break the
rotational symmetry of metasurface for the emergence of photonic SHE
in the momentum space~\cite{Liu2015}. The dielectric metasurfaces
are chosen as the examples to demonstrate our deduction. When the
phase retardances are fixed at $\pi$, the metasurfaces can be viewed
as the a spatially variant half-waveplate. It is well known that the
half-waveplate will convert the chirality of circularly polarized
beams, which lead to the evolution of states along the longitudes of
Poincar\'{e} sphere. Thus the process will endow the light an
additional PB phase. From the Eq.~(\ref{GPII}), it can be deduced
that the PB phase equals half of the surface area enclosed be the
evolution route $C$, where $l$ and $m$ both equal zero for these
localized situations. Furthermore, the evolving longitudes are
decided by the orientation of optical axes $\phi$ of the
inhomogeneous half-waveplate and the sign of PB phase is decided by
the incident polarization state $\sigma$. By this way, a spatially
variant phase distribution can be induced by engineering the
orientations of local optical axes.

A metasurface with rotational symmetry is designed in our work, as
shown in Fig~\ref{split}(a). It is well known that the phase
variation across the beam would deflect the light
$\Delta\textbf{k}=\nabla\Phi$ ($\textbf{k}$ is the wave vector).
Note that the geometric phase gradient in the azimuthal direction is
proportional to the rotation rate of local optical axes
$\Omega=d\phi/dx$, and that it is spin-dependent in sign but equal
in magnitude. When we refer to a small part of the metasurface as
the input light isn't located at the center of the metasurface, the
phase gradient can be approximately regarded as in one dimension
($x$). Thus, the constructed geometric phase gradient can be related
to the rotation rate of local optical axes in the form:
\begin{equation}
\Delta_{x}\textbf{k}=\nabla_{x}\Phi=d\Phi/dx=\frac{\sigma_{\pm}2d\phi}{dx}=2\sigma_{\pm}\Omega\label{PBPG},
\end{equation}
where $\Phi$ is the induced PB phase. Thus, when the metasurface is
engineered with rotational symmetry, the induced PB phase gradient
would deflect the light along the azimuth dimension.

However, the deflections of different spin components are opposite
as the PB phase is spin-dependent. Therefore, the wave vectors of
the two spin components are still interlaced after passing the
metasurface, thus no spin-dependent splitting in position space
occurs. By breaking the rotational symmetry of the metasurface, the
kept part of the metasurface can still deflect the light, but the
wave vectors cannot be still interlaced after propagation, as shown
in Fig~\ref{split}(b). By this way, we observed the photonic SHE in
our experiments. Furthermore, we reported the observation of
photonic SHE near the phase singularity at dielectric
metasurfaces~\cite{Li2015}. The rotationally arranged structures
(structure parameter $q$) are applied in this experiment. For a
linearly polarized incident vortex beam (topological charge $l$),
the spin-orbit coupling of light in the metasurface will lead to the
appearance of the helical phase with topological charge $\pm2q$ for
the opposite spin components respectively. Therefore, the
superposition of the induced PB phase and the original vortex phase
$l\pm2q$ will eliminate the phase singularity of one spin component
($l-2q = 0$ or $l+2q = 0$), while enhances the phase singularity of
the other. By switching the signs of the topological charges of the
input beam, the spin-dependent splitting pattern can be manipulated
effectively.

Moreover, there is still not an effective method to generate all the
states on higher-order Poincar\'{e} sphere and realize the state
evolution. Most efforts have been made to obtain the radial and
azimuthal polarized beams, but the other states are seldom referred
to. We provided two feasible schemes to fill this blank. The first
one is based on the superposition of two orthogonal circularly
polarized vortex beams, where the chirality of the two vortex beams
are also opposite. A modified Mach-Zender interferometer is applied
to make the superposition and finally construct the arbitrary
cylindrical vector beams on the higher-order Poincar\'{e}
sphere~\cite{Chen2014}. For more simplicity, a dielectric
metasurface was designed to realize this purpose~\cite{Liu2014}. By
just adjusting the incident fundamental polarization state, we can
access arbitrary states on higher-order Poincar\'{e} sphere.
Furthermore, by simply rotating a polarizer or a quarter waveplate
in our setup, the produced state will evolve along the latitude or
longitude of the higher-order Poincar\'{e} sphere. Moreover, a
versatile scheme based on two cascaded metasurfaces is proposed to
acquire arbitrary linearly polarized vector vortex states, which are
the particular points of the higher-order Poincar\'{e} sphere and
the hybrid-order Poincar\'{e} sphere~\cite{Yi2014}.

\begin{figure}
 \centerline{\includegraphics[width=17cm]{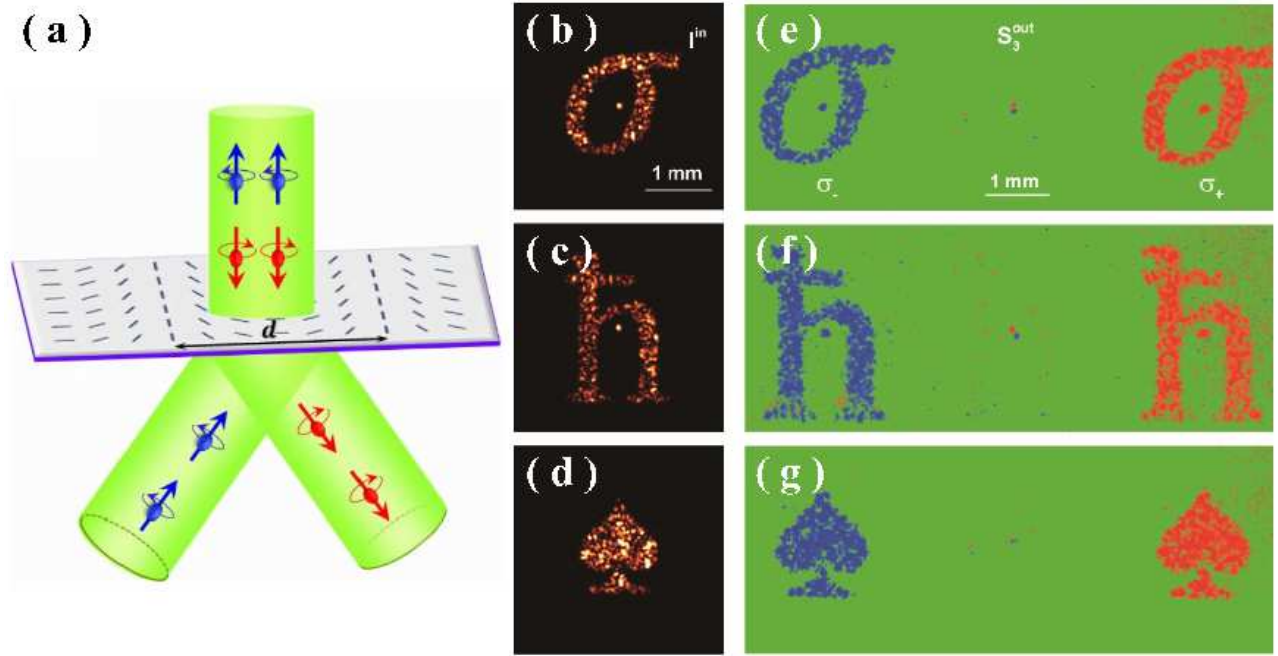}}
 \caption{(a) Schematic illustration of spin-dependent splitting. The
small balls with arrows represent the left- (red) and right-handed
(blue) photons, respectively. The metasurface reverses the chirality
of incident photons and steers normally incident photons with
opposite handedness to two directions, due to space-variant PBP
varying in one dimension. (b)-(d) Intensity patterns (middle column)
of three typical linear polarization beams (two spin-dependent
letters $\sigma$ and $\hbar$ and one symbol $\spadesuit$ in the
order from top to bottom) before the dielectric metasurface. (e)-(g)
The corresponding spin-dependent splitting patterns after the
dielectric metasurface are discriminated by the normalized Stokes
parameter $S_{3}$ (right column).} \label{pbs}
 \end{figure}

\section{SPIN-PHOTONIC DEVICES}

In spite of the huge efforts devoted in the recent years, the
practical application of metal metamaterials/metasurfaces is still
restricted, as the limited high device efficiencies caused by the
Ohmic losses in the metal and the limited scattering cross sections
of the micro-elements~\cite{Sun2012b,Pors2013}. However, owing to
the comparatively small losses in the optical range, dielectric
metasurfaces provide the possibilities to realize the practically
high quality transmission-type optical devices, which is desperate
for the integrate optics~\cite{Lin2014,Hakobyan2014,West2014}. Here,
we report a giant photonic SHE generater, a spin-dependent beam
splitter, and a spin filter based on the dielectric metasurfaces.
The transparent dielectric materials is the guarantee of high
transmit efficiency, and the orientation related geometric PB phase
design makes it function-customizable for these dielectric
metasurfaces.

\begin{figure}
\centerline{\includegraphics[width=17cm]{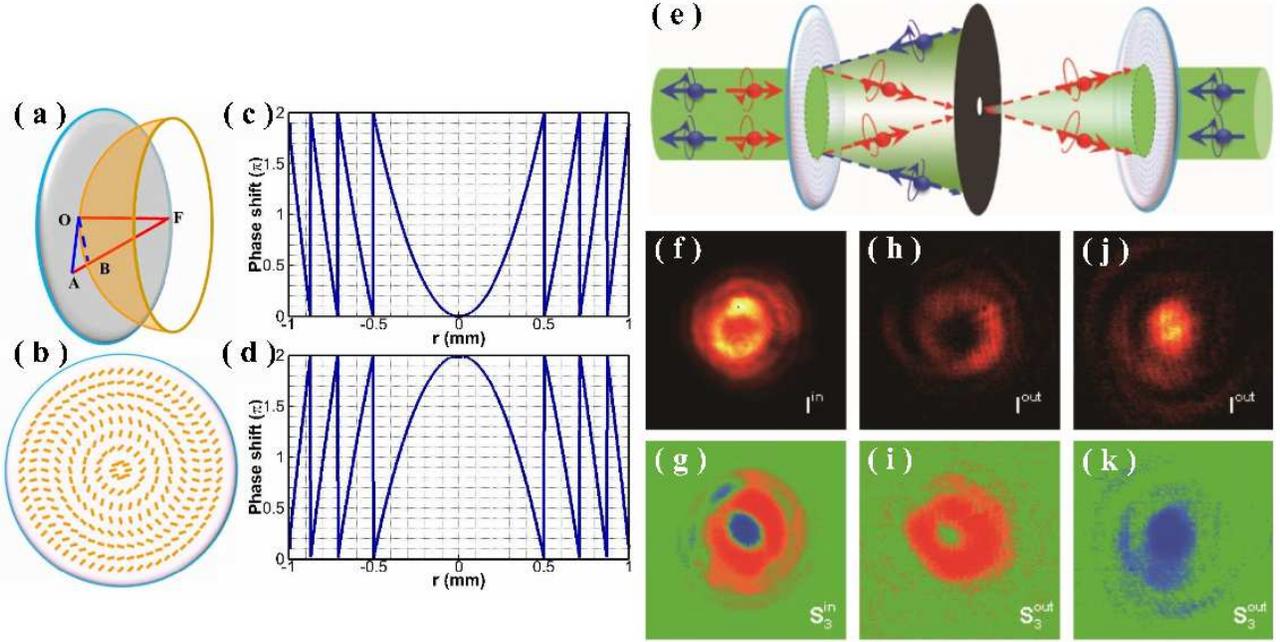}}
\caption{(a) The schematic illustration of a metasurface lens with
focal distance $f$ and the theoretical hyperboloidal equiphase
surface of this focusing lens. (b) The optical axis distribution of
a left-handed circular polarized plane wave focusing lens. (c)-(d)
The phase shift introduced by the same metasurface from the
identical incident plane, for left-handed circular polarization and
right-handed circular polarization beam incidence, respectively. (e)
Schematic setup for the photonic spin filter. The small balls with
arrows represent the left- (red) and right-handed (blue) photons,
respectively. The first metasurface reverses the chirality of
incident photons and results in the focusing of a selected spin
component, a diaphragm located at co-focus plane of the two
metasurfaces blocks the divergent spin component but releases the
selected one. Finally, another metasurface recovers the original
spin state and propagation direction of the emerging photons. (f)
and (g) show the intensity and normalized Stokes parameter $S_{3}$
of incident beam with the superposition of right-handed circularly
polarized plane Gaussian beam and left-handed circularly polarized
Laguerre-Gaussian beam with topological charge $1$. (h) and (i) are
respectively the intensity distribution and normalized Stokes
parameter $S_{3}$ of the emerging beam of right-handed filter
(blocking the right-hand photons and releasing the left-hand
photons). (j) and (k) are the contrary results for the left-handed
filter (blocking the right-hand photons and releasing the left-hand
photons).}\label{filter}
\end{figure}

As the photonic counterpart of electronic SHE, photonic SHE is a
rising star in recent years with high expectations in the precise
metrology, light manipulation, quantum communication, and quantum
computation~\cite{Yin2013,Shitrit2013,Hosten2008,Li2013,Zhou2012a,Zhou2012b,Zhou2013}.
Actually, this effect is the outcome of the mutual influence between
the photon spin (polarization) and the trajectory (orbital angular
momentum) of light-beam propagation, i.e.,
SOI~\cite{Onoda2004,Bliokh2008b,Bliokh2008a,Bliokh2006}. Thus the
photonic SHE is sensitive to the external disturbances, including
the material permittivity, dynamic shift, and ambient temperature.
Nonetheless, it is difficult to retrieve this variation by measuring
the photonic SHE. There is an exceptional difficult to detect this
effect directly, as the original SOI is generally very weak. A
precise but complex method dubbed quantum weak measurement is
developed in the past decade to detect the tiny spin-dependent shift
which is the manifestation of photonic
SHE~\cite{Hosten2008,Qin2009,Luo2011}. Meanwhile, it is also
profound to find a controllable and directly measurable photonic SHE
for the practical applications. We reported a giant photonic SHE
generater in momentum space based on the dielectric
metasurface~\cite{Ling2015}. In our scenario, the PB phase gradient
induces a constant momentum shift, i.e., a spin-dependent angular
deflection. The angular deflection in real-space is well-known that
increases linearly with the propagation distance. In addition, this
angular deflection is proportional to the rotation rate of the
metamaterial, and that the direction of spin shift can be reversed
by changing the sign of rotation rate. While the rotation rate can
be tuned by modulating the structural geometry of the metamaterial.
Based on the Eq.~(\ref{PBPG}), we get the splitting of light:
\begin{equation}
\Delta{x}=(\Delta{\mathbf{k}}/\mathbf{k}_{0})z=(2\sigma_{\pm}\Omega/\mathbf{k}_{0})z\label{PSHE},
\end{equation}
where $\mathbf{k}_{0}=2\pi/\lambda_{0}$, $\lambda_{0}$ is the
wavelength of incident light, $z$ is the propagation distance. In
compare with the photonic SHE that occurs in the interfacial
reflection and refraction, which exhibits a constant beam shift
limited to a fraction of the wavelength, the spin-dependent shift
induced in our scheme is sufficiently large to allow for direct
measurement. Therefore, such a structured metamaterial offers an
additional degree of freedom for the manipulation of the photonic
SHE, especially in the effort to obtain a giant photonic SHE.

Polarizing beam splitters which can separate the two orthogonal
polarizations of light beam into different propagation directions,
are widely used in optical communications, imaging systems, and
optical recording. Conventional polarizing beam splitters made by
naturally anisotropic materials require a large thickness to
generate the enough walk-off distance between the two orthogonal
polarizations, owing to the intrinsically small birefringence.
However, it is still a challenge to realize the circular
polarization beam splitter due to the lack of natural materials with
sufficient circular birefringence. Recently, benefitting from the
prosperity of artificial materials in the nearest researches,
including metamaterials, plasmonic metasurfaces, and the materials
designed in transform optics, many novel schemes are proposed for
the realization of beam splitters with special characteristics.
Here, we demonstrate a circular polarization beam splitter base on
the PB phase dielectric metasurface. As illustrated in
Fig.~\ref{pbs}, the metasurface separates the normally incident
light to two distinct parts with opposite helicities due to the
engineered PB phase gradient. The same as what we have mentioned
above, the linearly rotated local optical axes in dielectric
metasurfaces will introduce a spin dependent phase gradient at the
beam transverse section. Therefore, circularly polarized beams will
be separated according to their polarization. By this way, the light
beam encoded with arbitrary intensity patterns can be separated
after passing through the metasurface spin splitter, and that the
splitting angle can be adjusted according to the Eq.~(\ref{PSHE}).
Figures~\ref{pbs}(e)-\ref{pbs}(g) exhibit the observed splitting
results in out experiment.

Spin of photon provides a new degree of freedom to control light,
and enables many new spin-based applications. The spin filter which
screens out the pure spin photons is essential for the application
of this degree of freedom, similar to those in the spintronics. We
reports the experimental demonstration of the photon spin filter
based on two cascaded metasurfaces which is constructed with the
varying local optical axes in the radial direction. Firstly, a
dielectric metasurface lens is designed as shown in
Fig.~\ref{filter}(a). By locally adjusting the optical axis
orientations, as the Fig.~\ref{filter}(b) showing, the induced PB
phases for the opposite spin components are distributed as the
Fig.~\ref{filter}(c) and Fig.~\ref{filter}(c) showing. It is clear
that the PB phases will focus the left-hand circularly polarized
component while diverge the right-hand component. Therefore, based
on the cascaded metasurface lenses and a diaphragm, we obtain a
optical spin filter. As shown in Fig.~\ref{filter}(e), the first
metasurface reverses the chirality of incident photons and imposes a
PB phase on the incident light, thus resulting in a spin-dependent
expansion or focus for the opposite spin components of light. The
focused photons then pass through a diaphragm and the unwanted spin
component is blocked. Another metasurface is employed to recover the
spin state and propagation direction of the selected component.
Figures~\ref{filter}(f)-~\ref{filter}(k) exhibit the ability of this
spin-filter, normally incident photons with right-handed
(left-handed) spin are almost totally blocked, whereas normally
incident photons with lefthanded (right-handed) spin are almost
totally transmitted. To distinguish the opposite spin components, a
fundamental Gaussian mode is endowed to the left-hand spin
component, and a Laguerre-Gaussian mode is endowed to the right-hand
component. The passing spin component can be reversed by engineering
the metasurface with an opposite spatial rotation rate.

\section{CONCLUSIONS}
In summary, we have reviewed our recent works on spin photonics and
Spin-photonic devices with dielectric metasurfaces. The coupling
between a photon's spin state with its orbit degree of freedom is
known as the spin-orbit interaction, which contributes to two types
of geometric phase: the Rytov-Vladimirskii-Berry phase and the
Pancharatnam-Berry phase. The former is associated to the evolution
of propagation direction of light and the latter to the variation of
polarization state. It is feasible to manipulate the polarization
and propagation rule of light by designing the geometric phase.
However, the majority of traditional plasmonic metasurfaces are
based on the RVB phase or on the PB phase but being restricted to
near field. Dielectric metasurfaces fabricated of transparent
materials constitute a class of metasurface with considerable
flexibility in the manipulation of geometric phase and with high
transmission efficiency. We present here some of our works on
applying the dielectric metasurfaces to realize the evolution of
polarization states on Poincar\'{e} sphere and the manipulation of
propagation rule of light, moreover, we show that the dielectric
metasurfaces can develop novel spin-photonic devices such as spin
filter and spin-dependent beam splitter. It is the high transmission
efficiency of dielectric metasurfaces makes it possible to utilizing
the spin-dependent effect based on PB phase in optical far field,
which would substantially facilitate the development of
spin-dependent devices in photonics.

\acknowledgments  
This research was partially supported by the National Natural
Science Foundation of China (Grants Nos. 11274106 and 11474089).

\end{document}